\documentclass[11pt]{article}
\usepackage{graphicx}
\usepackage{amsmath}
\usepackage{subfigure}
\usepackage{graphicx}
\usepackage{amsmath}
\usepackage{amssymb}
\usepackage{enumitem}
\usepackage{booktabs}
\usepackage{enumitem}
\usepackage{cite}
\usepackage{color}
\usepackage[colorlinks]{hyperref}

\begin{document}
\begin{center}
{\Large\bf Thermodynamic Stability of a Viscous Variable Generalised Chaplygin Gas }\\[8 mm]
D. Panigrahi\footnote{ Netaji Nagar Day College, Kolkata 700092, India
\emph{and also} Relativity and Cosmology Research Centre, Jadavpur
University, Kolkata - 700032, India , e-mail:
dibyendupanigrahi@yahoo.co.in },
K. Kanjilal\footnote{Netaji Nagar Day College, Kolkata 700092, India, e-mail : kkanjilal92@gmail.com}
and Saibal Ray \footnote{Centre for Cosmology, Astrophysics and Space Science (CCASS), GLA University, Mathura 281406, Uttar Pradesh, India,
e-mail : saibal.ray@gla.ac.in}
\vspace{10mm}

\end{center}
\begin{abstract}
We investigate the cosmological dynamics and thermodynamic properties of a viscous variable generalized Chaplygin gas (VVGCG) model in a spatially flat Friedmann--Lema\^{\i}tre--Robertson--Walker universe, described by the equation of state $P=-B/\rho^\alpha$, where $B=B_0V^{n/3}$, with $B_0$ and $n$ constants and $V$ represents the volume of the fluid. The model includes a bulk-viscous pressure $\Pi=-3\xi H$, with $\xi=\xi_0\rho^{1/2}$. Unlike the usual assumption of a non-negative bulk-viscosity coefficient in linear irreversible thermodynamics, we find that the physically viable branch of the model corresponds to $\xi_0<0$. The resulting positive viscous pressure dominates in the early universe, giving rise to decelerated expansion, while the effective pressure subsequently changes sign at a finite volume $V_c$. The transition to acceleration occurs at a larger volume $V_f$, providing a natural mechanism for the deceleration-to-acceleration transition. At large volume, the equation-of-state parameter approaches a phantom-like asymptotic behaviour for $n<0$. A detailed thermodynamic analysis of the thermal and caloric equations of state, entropy, heat capacity, and isothermal and adiabatic stability establishes that the stability criteria and $c_V>0$ are simultaneously satisfied for $n<0$ and $\xi_0<0$. The entropy vanishes as $T\rightarrow0$, in accordance with the third law of thermodynamics, while the universe possesses a finite minimum volume and a corresponding maximum temperature. We further verify the thermodynamic consistency of the model using the fundamental thermodynamic identity. The general formulation of the model also contains several existing Chaplygin-gas models as special cases. Constraints from the Hubble-57 data yield the best-fit values $\alpha=0.0024$, $n=-0.0182$, and $\xi_0=-0.0015$, which lie within the thermodynamically stable parameter region. These results demonstrate that the VVGCG model provides a consistent description connecting the early decelerating universe with a late-time phantom-like accelerating phase, while remaining thermodynamically stable throughout the evolution.
\end{abstract}

   ~~~~KEYWORDS : cosmology; Chaplygin gas;thermodynamics; bulk viscosity\\
   \vspace{-0.3cm }

~~\hspace{0.2 cm} PACS :   98.80.-k,98.80.Es,95.30.Tg,05.70.Ce

\section{ Introduction}
The discovery of the universe's late-time accelerated expansion \cite{rei,nop, aman} presents a profound challenge to modern physics and cosmology. Data from multiple NASA missions \cite{spe1,spe2} show that ordinary baryonic matter---which forms stars and galaxies---accounts for less than $5\%$ of the total energy content of the cosmos. Additionally, independent cosmological probes indicate that roughly $73\%$ of the total energy density consists of an unknown component called \emph{dark energy}, while approximately $22\%$ is made up of non-baryonic cold dark matter that clusters gravitationally yet remains directly undetected.

Commonly attributed to dark energy---a gravitationally repulsive component of the cosmic fluid---observed cosmic acceleration remains one of modern cosmology's most perplexing open problems despite intensive theoretical and observational efforts. Although numerous theoretical models have been proposed to address this phenomenon (comprehensive reviews can be found in Refs.~\cite{neu,sah}), different variants of the Chaplygin gas have emerged as particularly promising candidates. However, researchers have repeatedly emphasized that extending the Chaplygin gas equation of state~\cite{kam} merely by adding free parameters~\cite{ben,ud,guo1, dp} without clear physical motivation does not yield a satisfactory model. While such generalizations might enhance the fit to observational data, ensuring the underlying fluid's physical consistency is equally crucial. Specifically, a viable dark energy candidate must constitute a closed thermodynamic system and satisfy fundamental stability criteria, which in turn place strict constraints on the model's allowed parameter space.

Chaplygin gas cosmology \cite{kam,ben} offers an appealing framework that unifies dark matter and dark energy into a single exotic fluid. Although such an equation of state is inapplicable in the primordial universe \cite{ben}, it exhibits a notable late-time feature: the fluid behaves like pressure-less dust at early epochs and evolves asymptotically toward a cosmological constant-dominated phase. The generalized Chaplygin gas (GCG) \cite{ben} equation of state is given by
\begin{equation}\label{eq:eq1}
P = -\frac{B}{\rho^{\alpha}},
\end{equation}
where $P$ and $\rho$ denote the pressure and energy density of the fluid, respectively, and $B$ is a positive constant.

A variable generalized Chaplygin gas (VGCG) model was recently proposed \cite{guo2} and constrained using Union supernova data and baryon acoustic oscillation (BAO) measurements \cite{lu1}, with the parameter $B$ assumed to depend on the scale factor of the Friedmann--Robertson--Walker universe. In this work, we adopt the parametrization $B = B_{0} V^{-n/3}$, where $V$ represents the volume of the fluid, $B_{0}$ is a constant, and $n$ is a free parameter. When $n=0$, the VGCG equation of state naturally reduces to that of the standard generalized Chaplygin gas \cite{ben}.

Observational constraints on the parameter $n$ exhibit considerable uncertainty. While Lu \cite{lu1} obtained $n = 0.75$ and $\alpha = 1.53$, a subsequent analysis with $\alpha = 1$ yielded $n = 1.30^{+0.46}_{-0.07}$ ($1\sigma$) using combined SNe Ia and BAO data \cite{lu2}. Interestingly, when considering only SNe Ia Union data, both positive and negative values of $n$ are allowed. Similarly, Sethi \emph{et al.}~\cite{set} found that for $\alpha = 1$, the best-fit values of $n$ lie within the range $(-1.3,\,2.6)$ based on WMAP and SNe Ia observations, meaning current data do not exclude either sign. In the present work, we obtain a negative value for $n$, as summarized in Table~\ref{tab:contour}.

The sign of $n$ carries significant physical implications. Positive values foster a quiescent cosmological evolution that avoids a big-rip singularity, whereas negative values correspond to a phantom-like Chaplygin gas scenario in which the dark energy density increases with cosmic time. Furthermore, recent observational analyses suggest that a strongly negative equation of state parameter ($w \leq -1$) may provide a better fit to cosmological data \cite{guo1,guo2,alen}. Although extensive literature exists on fitting Chaplygin gas models to observational data, agreement with observations alone is insufficient; the thermodynamic viability of the fluid must also be carefully examined. Santos \emph{et al.}~\cite{san1} demonstrated that the generalized Chaplygin gas model is thermodynamically stable, and Panigrahi et al \cite{dp1, dp2, dp3, dp4} showed that the VGCG model remains stable for $n < 0$. Building upon these foundations, the primary motivation of the present work is to investigate how the inclusion of bulk viscosity affects the thermodynamic stability conditions of the variable generalized Chaplygin gas. In this context, it is relevant to note that if the Chaplygin gas is regarded as an effective medium with internal degrees of freedom, particle interactions, relaxation processes, or a departure from local thermodynamic equilibrium, cosmic expansion can give rise to an effective bulk-viscous pressure.

Within an expanding Friedmann--Lema\^{\i}tre--Robertson--Walker (FLRW) spacetime, bulk viscosity modifies the effective pressure of a cosmic fluid via the bulk-viscous pressure, typically expressed as $\Pi = -3\xi H$, where $\xi$ denotes the bulk-viscosity coefficient and $H$ is the Hubble parameter. In linear, near-equilibrium thermodynamics, Onsager's reciprocal relations \cite{ons1,ons2} together with the Second Law require a non-negative rate of entropy production per unit volume ($\sigma \geq 0$), which generally implies $k \geq 0$ \cite{ons1,ons2}. However, in non-linear, far-from-equilibrium systems with continuous energy input, Prigogine \emph{et al.}~\cite{pri} showed that localized effective transport coefficients may become negative without violating global thermodynamic laws. This provides a physical motivation for considering an effective negative bulk-viscosity coefficient ($\xi < 0$) in cosmology. In this regime, $\Pi > 0$, meaning the viscous contribution reduces the overall negative pressure and can favor a decelerating phase at early times, smoothing the transition to accelerated expansion.

The interplay between bulk viscosity and Chaplygin-type fluids has been explored extensively in various broader contexts \cite{saa, bre1,bre2}. For instance, viscous generalized Chaplygin gas (VMCG) and modified Chaplygin gas (VMCCG) frameworks integrate bulk viscosity alongside time-dependent parameters \cite{bre2, ama}, whereas variable Chaplygin gas (VCG) models have also been investigated in non-flat FLRW spacetimes \cite{xu, bar}. Motivated by these developments, we examine the interplay between bulk viscosity and the Chaplygin-gas component within a spatially flat FRW universe. Following Ref.~\cite{tha}, we adopt the phenomenological form $\xi = \xi_0 \rho^{1/2}$~\cite{bar} for the bulk-viscosity coefficient, where $\xi_0$ is a constant and $\rho$ denotes the fluid's energy density.

Our analysis reveals that the thermodynamic stability of the viscous variable generalized Chaplygin gas (VVGCG) model imposes simultaneous constraints on its parameters, requiring both a negative volume-evolution exponent ($n < 0$) and a negative bulk-viscosity amplitude ($\xi_0 < 0$). For the parametrization $\xi = \xi_0\rho^{1/2}$, a negative $\xi_0$ yields a positive viscous pressure ($\Pi > 0$), which offsets the negative pressure of the Chaplygin component. This stability requirement thus serves as an independent physical restriction on the viscous sector, complementing constraints derived from cosmological observations.

The paper is organized as follows. In Sec.~\ref{ee}, we derive the expression for the energy density using the thermodynamic relations involving internal energy and entropy. Section~\ref{tb} is devoted to the evolution of the pressure, caloric equation of state, deceleration parameter, squared sound speed, and other relevant quantities with the volume for the variable generalised Chaplygin gas with bulk viscosity, examining how bulk viscosity influences cosmic evolution for both positive and negative values of $\xi_0$. In Sec.~\ref{tsc}, we investigate the thermodynamic stability of the model and show that the stability conditions $(\partial P_{\rm eff}/\partial V)_{S} < 0$, $(\partial P_{\rm eff}/\partial V)_{T} < 0$, and $c_V > 0$ are satisfied for $n < 0$ and $\xi_0 < 0$. Section~\ref{ctr} is devoted to checking the derived results using thermodynamic identities as an additional consistency test. In Sec.~\ref{oc}, we constrain the model parameters using the Hubble--57 dataset through a standard $\chi^2$-minimization procedure, yielding negative values of both $n$ and $\xi_0$ at the minimum value $\chi^2_{\min}$, in agreement with the theoretical thermodynamic stability requirements. Finally, Sec.~\ref{dis} summarizes the main results and presents the concluding remarks. \\

\section{Energy equation }\label{ee}
The line element for a spatially flat FLRW spacetime is given by

\begin{equation}\label{eq:eq2}
ds^2 = dt^2 - a^2(t) \left(dr^2 + r^2 d \theta^2 + r^2 sin^2 \theta d \phi^2 \right),
\end{equation}
where $a(t)$ is the scale factor. Now, the Einstein equations can be expressed as

\begin{equation}\label{eq:eq3}
R_{\mu \nu} - \frac{1}{2} R g_{\mu \nu} = \frac{8 \pi G}{c^4} T_{\mu \nu},
\end{equation}
Throughout the analysis, we adopt the units $c=1$ and $8\pi G=1$. We consider the viscous variable generalized Chaplygin gas (VVGCG) as the dark-energy component of the Universe. For a bulk-viscous fluid, the energy-momentum tensor is given by

\begin{equation}\label{eq:e4}
T_{\mu \nu} = \left(\rho + P_{\rm eff} \right)u_{\mu}u_{\nu} -  P_{\rm eff} g_{\mu \nu},
\end{equation}
Here, $u^{\mu}$ represents the four-velocity vector, while $\rho$ denotes the energy density. Now we consider the equation of state for the variable generalized Chaplygin gas (VGCG)~\cite{guo2}, as given in Eq.~\eqref{eq:eq1}. In this model, $B=B_{0}V^{-n/3}$, where $B_{0}>0$ and $n$ is an arbitrary constant. The quantities $P$ and $\rho$ denote the pressure and energy density of the fluid, respectively. The total effective pressure, including the contribution from bulk viscosity, can therefore be expressed as

\begin{equation}\label{eq:eq5}
P_{\rm eff} = P + \pi =   - \frac{B_{0}V^{-\frac{n}{3}}} {\rho^{\alpha} } - \sqrt{3} \xi_0 \rho
\end{equation}
The energy density of the viscous variable
generalised Chaplygin gas (VVGCG) is then determined from the corresponding energy-density relation as

\begin{equation}\label{eq:eq6}
\rho_{\rm vvgcg} = \frac{U}{V},
\end{equation}
Here, $U$ and $V$ denote the internal energy and the volume occupied by the fluid, respectively. We now derive the internal energy $U$ and pressure
$P_{\rm eff}$ of the variable generalised Chaplygin gas as functions of the entropy $S$ and volume $V$. From the fundamental relations of thermodynamics, we have
\begin{equation} \label{eq:eq7}
\left(\frac{\partial U }{\partial V}\right)_{S}= - P_{\rm eff}.
\end{equation}
Using Eqs.~\eqref{eq:eq5} and \eqref{eq:eq7}, we can determine the internal energy $U$ and pressure $P$ of the viscous variable generalised Chaplygin gas in terms of the entropy $S$ and volume $V$. Accordingly, the above thermodynamic relations yield

\begin{equation}\label{eq:eq8}
\left(\frac{\partial U }{\partial V}\right)_{S}=
B_{0}V^{-\frac{n}{3}} \frac{V^{\alpha}}{U^{\alpha}} + \sqrt{3}\xi_0 \frac{U}{V},
\end{equation}
which, upon integration, yields

\begin{equation} \label{eq:eq9}
U = \left[ \frac{3B_{0}(1+\alpha)V^{\frac{3(1+\alpha)-n}{3}}}
{3(1 - \sqrt{3}\xi_0)(1+\alpha)-n} + \frac{c}{V^{ - \sqrt{3}\xi_0(1+\alpha)}}
\right]^{\frac{1}{1+\alpha}}.
\end{equation}
Here, $c$ denotes the integration constant, which may either be a universal constant or a function of the entropy $S$ alone. We can therefore recast the above equation in the following form

\begin{equation}\label{eq:eq10}
U  = \left[\frac{B_{0}(1+\alpha)V^{-\frac{n}{3}}}{N}\right]^{\frac{1}{1+\alpha}}
V \left[1+
\left(\frac{\epsilon}{V}\right)^N\right]^{\frac{1}{1+\alpha}},
\end{equation}
where
\begin{equation}\label{eq:eq10a}
N=(1-\sqrt{3}\xi_0)(1+\alpha)-\frac{n}{3}.
\end{equation}
It follows from Eq.~\eqref{eq:eq10} that $N$ must be positive for $U$ to remain real. For $0<\alpha<1$, this requires
$(1-\sqrt{3}\xi_0)(1+\alpha)>\frac{n}{3}.$ A sufficient condition is $\xi_0<0$ and $n<0$, for which $N>0$ is automatically satisfied.  Alternatively, for $0<\xi_0<1/\sqrt{3}$, the same requirement is fulfilled when $n<0$. Now we consider

\begin{equation}\label{eq:eq11}
\epsilon =
\left[\frac{3(1 - \sqrt{3}\xi_0)(1+\alpha)-n}{3B_{0}(1+\alpha)}~c\right]^{\frac{1}{N}}
= \left[\frac{Nc}{B_{0}(1+\alpha)}\right]^{\frac{1}{N}},
\end{equation}
which has the dimensions of volume. Substituting this result into Eq.~\eqref{eq:eq6}, we obtain the energy density $\rho$ of the VVGCG in terms of the entropy $S$ and volume $V$ as

\begin{subequations}\label{eq:eq12}
\begin{align}
\rho_{\rm vvgcg} &= \left[\frac{B_{0}(1+\alpha)V^{-\frac{n}{3}}}{N} + V ^{-N - \frac{n}{3}} c\right]^{\frac{1}{1+\alpha}}  \label{eq:eq12a}\\
&= \left[\frac{B_{0}(1+\alpha)V^{-\frac{n}{3}}}{N}\right]^{\frac{1}{1+\alpha}}
 \left[1+
\left(\frac{\epsilon}{V}\right)^N\right]^{\frac{1}{1+\alpha}}  \label{eq:eq12b}.
\end{align}
\end{subequations}
Based on the discussions presented above, we now derive the relevant physical quantities and examine their behaviour throughout the evolution.\\

\section{Thermodynamical behaviour}\label{tb}
\subsection{Pressure }\label{pr}

Using Eqs  \eqref{eq:eq5} and  \eqref{eq:eq12b} the pressure $P_{\rm eff}$ of the VVGCG  may also be determined as a function of
entropy $S$ and volume $V$  in the following form

\begin{eqnarray}\label{eq:eq13}
P_{\rm eff} =-\left[\frac{B_{0}(1+\alpha)V^{-\frac{n}{3}}}{N}\right]^{\frac{1}{1+\alpha}}
\left(\frac{N}{1+\alpha}\right) \frac{\left[1 + \frac{\sqrt{3}\xi_0(1+\alpha)}{N} \left\{1+
\left(\frac{\epsilon}{V}\right)^N \right \}\right]}{\left[1+
\left(\frac{\epsilon}{V}\right)^N\right]^{\frac{\alpha}{1+\alpha}}}.
\end{eqnarray}
The Eq.~\eqref{eq:eq13} provides a general expression for the effective pressure of the VVGCG fluid, whose signature depends on the magnitude and sign of $\xi_0$ and $n$, as well as on the volume $V$.
\begin{figure}[t]
    \centering
    \subfigure[\small\emph{$P_{\rm eff}$ is positive for small $V$ and negative for large $V$ for $n = - 0.0182$ \& $\xi_0 = - 0.0015$}]{
        \includegraphics[width=6 cm]{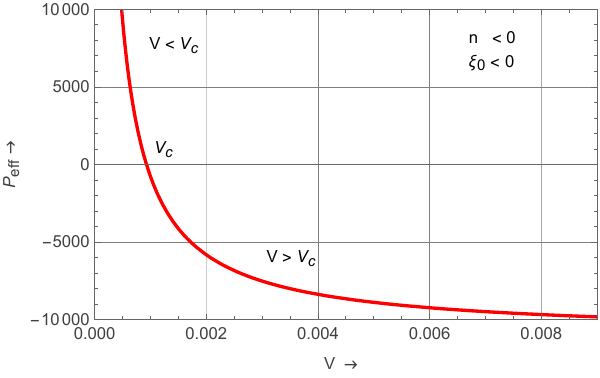}
        \label{pv1}
    }
    ~~~
    \subfigure[\small\emph{$P_{\rm eff}$ is negative throughout the evolution  for  $n = - 0.0182$ \& $\xi_0 =  0.0015$}]{
        \includegraphics[width=6 cm]{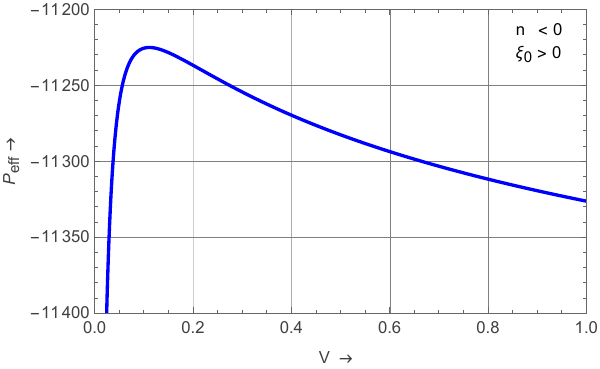}
        \label{pv2}
    }
    \caption[\small Optional caption for list of figures]{
        \small\emph{The variations of $P_{\rm eff}$ and $V$}
    }
    \label{pv}
\end{figure}
The parameter $\xi_0$ determines the contribution of bulk viscosity to the effective pressure. Since the viscous pressure is given by $\pi=-\sqrt{3}\xi_0\rho$, a positive $\xi_0$ produces a negative viscous pressure, $\pi<0$, which reinforces the negative pressure of the Chaplygin component. Consequently, for $\xi_0>0$, $P_{\rm eff}<0$ throughout the evolution and the fluid favours accelerated expansion. In contrast, for $\xi_0<0$, the viscous contribution is positive, $\pi>0$, and therefore opposes the negative pressure of the Chaplygin component. In this case, the relative contributions of the two terms determine the sign of $P_{\rm eff}$, which consequently depends on the volume $V$.

So for $\xi_0<0$, in the small-volume limit, $V\rightarrow0$, the term $(\epsilon/V)^N$ becomes dominant, resulting in $P_{\rm eff}>0$. As the volume increases, the effective pressure vanishes at a finite critical volume $V_c$ and subsequently becomes negative. Thus,
$ P_{\rm eff}>0\quad\text{for}\quad V<V_c$,
$ P_{\rm eff}=0\quad\text{for}\quad V=V_c$, and
$ P_{\rm eff}<0\quad\text{for}\quad V>V_c$. The critical volume, obtained by setting $P_{\rm eff}=0$, is given by  Eq. \eqref{eq:eq13} as
\begin{equation}\label{eq:eq13a}
V_c=\epsilon\left[ \frac{-3\sqrt{3}\xi_0(1+\alpha)} {3(1+\alpha)-n}\right]^{1/N}.
\end{equation}
For $\xi_0<0$, the numerator in Eq.~\eqref{eq:eq13a} is positive, while the denominator is positive provided $n < 0$ or $n<3(1+\alpha)$. In the present analysis, we impose the stronger condition $n<0$, which automatically satisfies this requirement. Hence, $V_c$ is real and finite.

The above behaviour demonstrates that the sign of $\xi_0$ plays a crucial role in determining the cosmic dynamics. For $\xi_0>0$, the viscous pressure is negative and reinforces the negative Chaplygin pressure, favouring accelerated expansion. For $\xi_0<0$, the positive viscous pressure can dominate at sufficiently small volumes, giving rise to a positive effective pressure and hence a decelerating phase. As the universe expands beyond $V_c$, the negative Chaplygin contribution becomes dominant, causing $P_{\rm eff}$ to change from positive to negative. Thus, $V_c$ should be interpreted specifically as the characteristic volume at which the effective pressure changes sign. The corresponding behaviour of $P_{\rm eff}$ as a function of $V$ is shown in Fig.~\ref{pv}. For the special case $\xi_0=0$, the viscous contribution vanishes, so $\Pi=0$ and $V_c=0$. Hence, there is no finite positive volume at which the effective pressure changes sign, and the pressure-transition behaviour associated with negative bulk viscosity is absent. The model then reduces to the non-viscous variable generalized Chaplygin gas case. We now examine the behaviour of the VVGCG fluid under some specific conditions.

\begin{enumerate}[label=(\roman*)]
\item  For $n=0$ \& $\xi_0 = 0$, the Eq. \eqref{eq:eq13}   reduces to
\begin{equation}\label{eq:eq14}
 P_{\rm eff} = - \frac{(B_{0})^{\frac{1}{1+\alpha}}}{\left \{1+\left(\frac{c}{B_0 V}\right)^{1+\alpha} \right \}^{\frac{\alpha}{1+\alpha} }},
\end{equation}
which, as expected, exhibits behaviour similar to that of the generalised Chaplygin gas (GCG) model~\cite{san1}.\\

\item  For $n=0$ \& $\xi_0  \neq 0$, we get the generalized Chaplygin gas (GCG) with bulk viscosity model, the pressure given by
\begin{eqnarray}\label{eq:eq15}
P_{\rm eff}  = - \left(\frac{B_0}{1 - \sqrt{3}k_0}\right)^{\frac{1}{1+\alpha}} \frac{\left[1+ \sqrt{3}\xi_0 \left\{ \frac{(1 - \sqrt{3}\xi_0) c}{B_0V} \right \}^{(1 - \sqrt{3}\xi_0)(1+\alpha)}\right] }{\left[1+\left \{\frac{(1 - \sqrt{3}\xi_0) c}{B_0V}  \right \}^{(1 - \sqrt{3}\xi_0)(1+\alpha)}\right]^{\frac{\alpha}{1+\alpha}}}.
 \end{eqnarray}
Thus, for either $\xi_0<0$ or $0<\xi_0<\frac{1}{\sqrt{3}}$, the effective pressure remains negative, i.e., $P_{\rm eff}<0$.\\

\item  For $\xi_0 = 0$, but $n \neq 0$, the model represents the Variable generalised Chaplygin gas (VCGC)~\cite{dp3} and Eq.~\eqref{eq:eq13} becomes
 \begin{equation}\label{eq:eq16}
 P_{\rm eff} = - \left(B_{0} V^{-\frac{n}{3}} \right)^{\frac{1}{1+\alpha}}
 \left[\frac{N}{(1+ \alpha) \left \{1 + \left(\frac{\epsilon}{V} \right)^{N}\right\}} \right]^{\frac{\alpha}{1+\alpha}}.
\end{equation}
which corresponds to the standard inviscid form where the viscous pressure correction vanishes identically.
Again, setting $\alpha = 1$ in Eq.~\eqref{eq:eq16}, we obtain
$ P_{\rm eff} = - \left[ \frac{N B_0 V^{-\frac{n}{3}}}{2 \left\{1 + \left(\frac{\epsilon}{V}\right)^N\right\}}\right]^{\frac{1}{2}}$, which is identical to that presented in an earlier article~\cite{dp3, dp4} when considering the Variable Chaplygin Gas (VCG) model.\\
\item For $\xi_0 \neq 0$, $n \neq 0$, and $\alpha = 1$, the model represents the viscous Variable Chaplygin gas with viscosity, and Eq.~\eqref{eq:eq13} becomes

\begin{equation}\label{eq:eq17}
 P_{\rm eff} = -  \left(\frac{ N B_{0} V^{-\frac{n}{3}}}{2\left \{1 + \left(\frac{\epsilon}{V} \right)^{N}\right\}} \right)^{\frac{1}{2}}
 \left[(1+ \frac{2 \sqrt{3}\xi_0} {N} \left \{1 + \left(\frac{\epsilon}{V} \right)^{N}\right\} \right].
\end{equation}
which is identical to the result obtained by Barman and Barman ~\cite{bar}.
\end{enumerate}

Thus, with the expansion of the universe, an initially decelerating phase begins to accelerate when the volume crosses the critical value $V_c$, governed by the effect of negative bulk viscosity. The existence of this critical volume is a direct consequence of the negative bulk-viscosity contribution. It is also worth noting that, for physically realistic values of the constants, both $V_c$ and $\epsilon$ are of the same order of magnitude. Hence, $\epsilon$ also represents a characteristic volume scale associated with the onset of the accelerating era. Consequently, $V\gg\epsilon$ corresponds to the large-volume limit, while $V\ll\epsilon$ corresponds to the small-volume limit. In what follows, we will see that this observation has significant cosmological implications.

\subsection{Caloric equation of state}\label{ceos}

Now using the expression   \eqref{eq:eq12b} and  \eqref{eq:eq13} we get the caloric
equation of state parameter
\begin{equation}\label{eq:eq18}
\mathcal{W }= \frac{P_{\rm eff}}{\rho_{\rm vvgcg}} = - \sqrt{3}\xi_0 - \left(\frac{N}{1+ \alpha}
\right) \frac{1}{1 + \left(\frac{\epsilon}{V}\right)^N } .
\end{equation}

\begin{figure}[t]
\begin{center}
   \includegraphics[width=8cm]{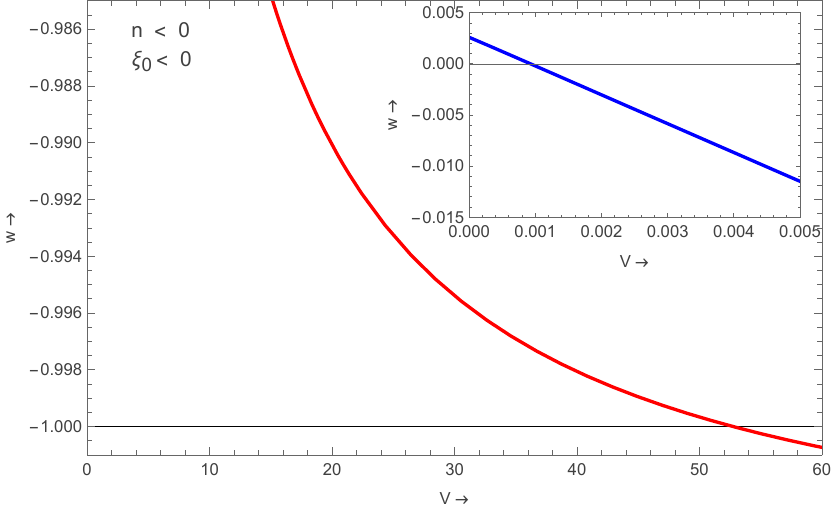}
     \caption{
  \small\emph{
  The variation of $\mathcal{W }$ and $V$  for  $n = - 0.0182$ \& $k_0 = - 0.0015$.
     }\label{wv}
    }
\end{center}
\end{figure}

We now study Eq.~\eqref{eq:eq18} for the extremal cases corresponding to different signs of $n$ and $\xi_0$, as follows: \\
\begin{enumerate}
\item For small volume, $V \ll \epsilon$, we get from equation  \eqref{eq:eq18} that

\begin{equation}\label{eq:eq19}
 P_{\rm eff} \approx  - \sqrt{3}\xi_0 \rho_{\rm vvgcg} .
 \end{equation}
This result further demonstrates the dependence of the effective pressure on $\xi_0$ at small volumes. For $\xi_0>0$, the viscous contribution $\pi<0$ lowers $P_{\rm eff}$ and can drive accelerated expansion, whereas for $\xi_0<0$, $\pi>0$ increases $P_{\rm eff}$ and can lead to a decelerating phase. For $\xi_0=0$, the viscous contribution vanishes, $\pi=0$, and the fluid reduces to the dust-dominated case with $P_{\rm eff}=0$, as discussed in Sec. \ref{pr}.

\item For large volume, $V \gg \epsilon$,  the Eq. \eqref{eq:eq18} reduces to

\begin{equation}\label{eq:eq20}
\mathcal{W }\approx -1 + \frac{n}{3(1+\alpha)}.
\end{equation}
Equation \eqref{eq:eq20} shows that an accelerating Universe requires either $0<n<3(1+\alpha)$ or $n<0$. So
  there are three possibilities for  $\mathcal{W }$ depending on the
  signature of $n$ as
\begin{enumerate} [label=(\roman*)]
\item $0 < n < 3(1+\alpha) $, $ \mathcal{W } > -1$, here the caloric equation of state results in a quiescence type  and big rip is avoided in this case,

\item $n = 0$, $\mathcal{W } = -1$, \emph{i.e.}, we get $\Lambda$CDM.

\item $n <0$, $\mathcal{W } < -1$ represents phantom like universe.
  This is compatible with recent observational results ~\cite{cim, kac}.
 The influence of $n$ is prominent in this case, while the effect of bulk viscosity is absent in the late Universe.
\end{enumerate}
\end{enumerate}
Initially, when the volume is sufficiently small, $\mathcal{W}>0$. As $V$ increases toward the critical volume $V_c$, $\mathcal{W}$ decreases to zero, reaching $\mathcal{W}=0$ at $V=V_c$. For $V>V_c$, $\mathcal{W}$ becomes negative. For $n>0$, in the large-volume limit, $-1<\mathcal{W}<0$, consistent with the result obtained earlier in the higher-dimensional case~\cite{dp6,dp7}. We do not consider the case $n>0$ further, since the thermodynamic stability conditions discussed later require $n<0$. For $\xi_0<0$ and $n<0$, Fig.~\ref{wv} shows the transition of $\mathcal{W}$ from positive to negative values with increasing volume.

\subsection{Deceleration parameter}\label{dep}

 Now using equation \eqref{eq:eq18} we calculate the deceleration parameter of the VVGCG fluid as
\begin{equation}\label{eq:eq21}
q = \frac{1}{2} + \frac{3}{2} \frac{P_{\rm eff}}{\rho_{\rm vvgcg}} = \frac{1}{2} -
\frac{3}{2}\left\{ \sqrt{3}\xi_0 + \left(\frac{N}{1+ \alpha} \right) \frac{1}{1
+ \left(\frac{\epsilon}{V}\right)^N }\right\} ,
\end{equation}

\begin{figure}[ht]
\begin{center}

   \includegraphics[width=8cm]{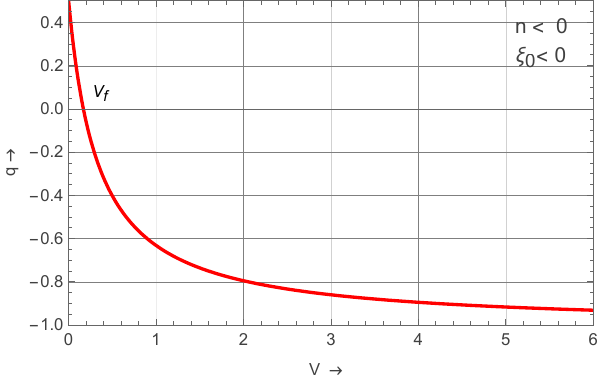}
     \caption{
  \small\emph{
  The variation of $q$ and $V$  for $n = - 0.0182$ \& $\xi_0 = - 0.0015$.
     }\label{qv}
    }
\end{center}
\end{figure}

For mathematical simplicity here also we discuss the extreme
cases.
\begin{enumerate}
\item  For small volume, $V \ll \epsilon$,
 it gives

\begin{equation}\label{eq:eq22}
q\approx  \frac{1}{2} - \frac{3}{2}\sqrt{3}\xi_0 ,
\end{equation}
\emph{i.e.}, for $\xi_0>\frac{1}{3\sqrt{3}}$, $q<0$, indicating accelerated expansion at small volumes. However, this regime is excluded by the positive-energy-density condition $\xi_0<\frac{1}{3\sqrt{3}}$. Thus, $q>0$ for the physically allowed range $\xi_0<\frac{1}{3\sqrt{3}}$, indicating decelerated expansion. In particular, for $\xi_0<0$, the positive viscous pressure reinforces the decelerating behaviour at sufficiently small volumes.\\

\item  For large volume, $V \gg \epsilon$, the equation \eqref{eq:eq21} reduces to
\begin{equation}\label{eq:eq23}
q\approx -1 + \frac{n}{2(1+\alpha)}.
\end{equation}
at large volumes, $q$ becomes negative, indicating an accelerating universe for $n < 0$ or $n < 2(1+\alpha)$.  Thus, the late-time accelerated expansion is governed predominantly by the parameter $n$, with the bulk-viscosity contribution becoming subdominant. For \emph{flip} in volume to occur the flip volume
($V_{f}$) becomes

\begin{equation}\label{eq:eq24}
V_{f} = \epsilon \left[\frac{(1  - 3\sqrt{3}\xi_0)(1+
\alpha)}{2(1+\alpha)-n}\right]^{\frac{1}{N}}.
\end{equation}
A brief analysis of the above equation shows that $V_f$ is real only for $0<n<2(1+\alpha)$ or $n<0$, provided either $0<\xi_0<\frac{1}{3\sqrt{3}}$ or $\xi_0<0$; otherwise, no \emph{flip} occurs. Thus, the conditions for a real $V_f$ and the occurrence of the \emph{flip} are $ 0<n<2(1+\alpha)\quad \text{or}\quad n<0$, with either $0<\xi_0<\frac{1}{3\sqrt{3}}$ or $\xi_0<0$. Interestingly, these conditions are consistent with the parameter restrictions considered above, namely $n<0$ and $\xi_0<0$. This behavior is also evident from Fig.~\ref{qv}, which shows the occurrence of the \emph{flip} for $n<0$ and $\xi_0<0$. We therefore focus on this parameter regime, since, as shown later, the VVGCG fluid is thermodynamically stable under these conditions. In this case, $V<V_f$ corresponds to a decelerating phase, whereas $V>V_f$ leads to an accelerating phase.
\end{enumerate}
It is worth noting that two characteristic volume scales, $V_c$ and $V_f$, arise in our analysis. These two scales have distinct physical meanings: $V_c$ corresponds to the vanishing of the effective pressure, $P_{\rm eff}=0$, whereas $V_f$ denotes the deceleration-to-acceleration transition. In a flat FRW cosmology, the condition for accelerated expansion is $\rho_{\rm vvgcg}+3P_{\rm eff}<0$. Therefore, negative pressure alone does not necessarily imply accelerated expansion; its magnitude must also be sufficiently large. For the present model, we obtain $V_c<V_f$, which also follows directly from Eqs.~\eqref{eq:eq17} and \eqref{eq:eq24}. At $V=V_f$, we find from Eq. \eqref{eq:eq21} that  $\rho_{\rm vvgcg}+3P_{\rm eff}=0$, as expected at the flip point. Thus, the effective pressure changes sign at $V_c$, while the actual transition from deceleration to acceleration occurs at the larger volume $V_f$.

\subsection{Square speed of sound}\label{vs}

Let us consider $v_s$ be the speed of sound,  then using Eq.  \eqref{eq:eq13} we can write

 \begin{equation}\label{eq:eq25}
v_{s}^2= \left(\frac {\partial P_{\rm eff}}{ \partial \rho{\rm vvgcg}} \right)_S =  - \sqrt{3}\xi_0 +
\frac{N \alpha}{ (1 + \alpha) \left \{1 + \left(\frac{\epsilon}{V}
\right)^N\right \} }  - \frac{n N}{n + (3N + n) \left(\frac{\epsilon}{V}\right)^N}.
 \end{equation}

 Since sound  speed should be   $0< \left(\frac {\partial P_{\rm eff}}{ \partial \rho_{\rm vvgcg}}
  \right)_S < 1$, now our analysis shows ,
\begin{enumerate}
\item  For small volume, \emph{i.e.}, at early universe, the Eq. \eqref{eq:eq25}
 leads to $v_s^2 \approx - \sqrt{3} \xi_0$ and it is positive for $\xi_0<0$.

\item For large volume, the Eq. \eqref{eq:eq25} reduces to

\begin{equation}\label{eq:eq26}
  v_{s}^2 =  - 1 + \frac{n}{3(1+ \alpha)}.
 \end{equation}

The Eq. \eqref{eq:eq26} does not depend on $\xi_0$, it depends on $n$ and
 $\alpha$ only. In what follows we shall see that from the thermodynamical stability condition   the value of
 $n < 0$,  leading to a phantom universe ~\cite{cim, kac}. Moreover the Eq.  \eqref{eq:eq26}
  gives an imaginary speed of sound for $\alpha > 0$, leading to a perturbative cosmology. One need not be too sceptic about it because  it favours structure formation~ \cite{fab}.
\end{enumerate}
It is worth comparing our results with the work of Y. S. Myung~\cite{my}, in which holographic dark energy, the Chaplygin gas, and the tachyon model with a constant potential were investigated. The squared speed of sound plays a crucial role in determining the stability of cosmological perturbations. Myung found that the squared speed of sound for holographic dark energy is always negative when the future event horizon is adopted as the infrared cutoff, whereas it remains non-negative for both the original Chaplygin gas and the tachyon model. In contrast, our VVGCG model exhibits a richer behaviour: the squared speed of sound can be either positive or negative, depending on the sign and magnitude of the parameter $n$ and $\xi_0$. This feature distinguishes the VVGCG model from the original Chaplygin gas and highlights the important role of the variable parameter $n$ and $\xi_0$ in determining the stability of perturbations. It is worth noting that the same results were obtained in our previous work~\cite{dp2}, where we investigated the thermodynamic stability of a variable modified Chaplygin gas.

However, our subsequent thermodynamic analysis shows that negative values of $n$ and $\xi_0$ are favoured for stability. This suggests that the VVGCG model may suffer from classical instability when $v_s^2<0$, similar to the holographic dark-energy scenario. Interestingly, our observational analysis also favours negative values of both $n$ and $\xi_0$, which are consistent with the thermodynamically stable parameter region.

\section{  Thermodynamic stability criteria}\label{tsc}

We now analyse the thermodynamic stability of the cosmic fluid throughout its evolution. For thermodynamic stability, the fluid must satisfy the following general conditions~\cite{land}:
\begin{enumerate}[label=(\roman*)]
\item The pressure must decrease with increasing volume under both adiabatic and isothermal processes, such that
$ \left(\frac{\partial P_{\rm eff}}{\partial V}\right)_{S}<0,
\qquad
\left(\frac{\partial P_{\rm eff}}{\partial V}\right)_{T}<0,
$

\item The specific heat at constant volume must remain positive, $c_V>0$.
\end{enumerate}
In the following subsections, we examine these stability criteria explicitly for the viscous variable generalised Chaplygin gas (VVGCG) model.

\subsection{Adiabatic Stability Condition}\label{asc}
We begin by analysing the stability of the fluid during adiabatic expansion. For mechanical stability, the pressure must decrease monotonically as the volume increases at fixed entropy, such that $ \left(\frac{\partial P_{\rm eff}}{\partial V}\right)_{S}<0$. If this condition is not satisfied, the fluid may become mechanically unstable during cosmic evolution. Using Eq.~\eqref{eq:eq13}, we obtain
\begin{eqnarray}\label{eq:eq27}
\left( \frac{\partial P_{\rm eff}}{\partial V}\right)_{S}
=\frac{P_{\rm eff}}{3V(1 + \alpha)}\frac{ -\sqrt{3}\xi_0(1+\alpha)  \left\{n \left[1 +
\left(\frac{\epsilon}{V} \right)^N \right]
 + 3N\left(\frac{\epsilon}{V} \right)^N \right \}
  + N \left\{\frac{3N \alpha \left(\frac{\epsilon}{V} \right)^N}{\left[
1+ \left(\frac{\epsilon}{V} \right)^N \right]} - n  \right\}  }{N+ \sqrt{3}\xi_0
(1+ \alpha)\left[1 +
\left(\frac{\epsilon}{V} \right)^N \right]}.~
\end{eqnarray}

\begin{figure}[t]
    \centering
    \subfigure[\small\emph{$\left(\frac{\partial P_{\rm eff}}{\partial V} \right)_S$ is negative through out the evolution for $n = - 0.0182$ \& $\xi_0 = - 0.0015$}]{
        \includegraphics[width=7cm]{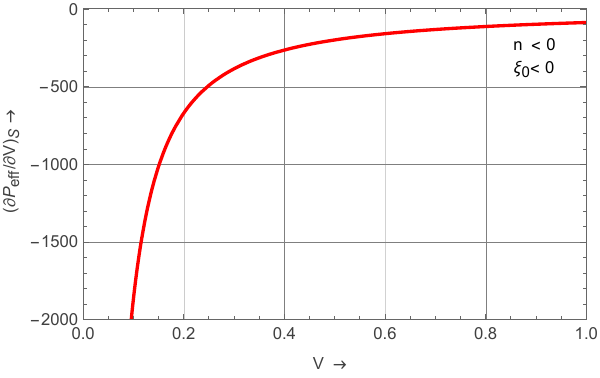}
        \label{dpv1}
    }
    ~~~
    \subfigure[\small\emph{$\left(\frac{\partial P_{\rm eff}}{\partial V} \right)_S$ is positive for small $V$ and negative for large $V$  for $n = - 0.0182$ \& $\xi_0 =  0.0015$}]{
        \includegraphics[width=7 cm]{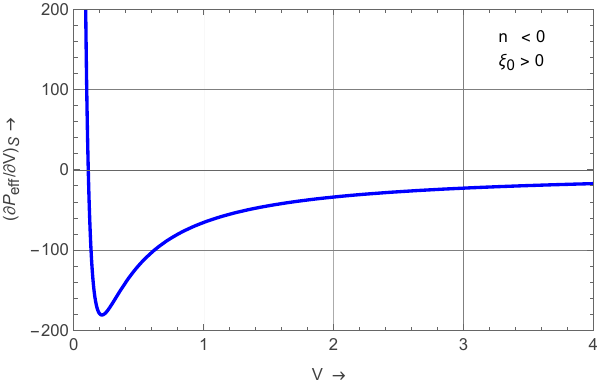}
        \label{dpv2}
    }
    \caption[\small Optional caption for list of figures]{
        \small\emph{The variations of $\left(\frac{\partial P_{\rm eff}}{\partial V} \right)_S$ and $V$}
    }
    \label{dpv}
\end{figure}

We now investigate the condition for
$\left(\frac{\partial P_{\rm eff}}{\partial V}\right)_{S}<0$.
Since the expression in Eq.~\eqref{eq:eq27} is rather involved, it is
difficult to draw a direct conclusion from the full expression.
Therefore, we analyse the stability condition separately in two
distinct volume regimes:

\begin{enumerate}
  \item  We begin by considering the small-volume regime, $V\ll\epsilon$. In this limit, Eq.~\eqref{eq:eq27} reduces to $ \left(\frac{\partial P_{\rm eff}}{\partial V}\right)_{S} \simeq -(1-\sqrt{3}\xi_0)\frac{P_{\rm eff}}{V}.$
From our earlier analysis in Eq. \eqref{eq:eq18}, the effective pressure at the early stage of evolution is approximately $ P_{\rm eff}\simeq -\sqrt{3}\xi_0\rho_{vvgcg}.$
Consequently, $ \left(\frac{\partial P_{\rm eff}}{\partial V}\right)_{S}
\simeq \sqrt{3}\xi_0(1-\sqrt{3}\xi_0)\frac{\rho_{\rm vvgcg}}{V}.$
Thus, in the small-volume limit, the thermodynamic stability is independent of $n$ and is determined primarily by $\xi_0$. For $\xi_0>0$, we have $
\left(\frac{\partial P_{\rm eff}}{\partial V}\right)_{S}>0,$
whereas for $\xi_0 < 0$, yielding  $(1 - \sqrt{3}\xi_0) > 0$,  resulting in $\left(\frac{\partial P_{\text{eff}}}{\partial V}\right)_S < 0$. which satisfies the thermodynamic stability condition. Hence, negative $k_0$ is favoured by the thermodynamic stability requirement in the small-volume regime.

\item In the large-volume regime, $V\gg\epsilon$, Eq.~\eqref{eq:eq27} takes the approximate form $ \left(\frac{\partial P_{\rm eff}}{\partial V}\right)_S
\simeq -\frac{nP_{\rm eff}}{3V(1+\alpha)}$. Since $P_{\rm eff}<0$ at late times, the thermodynamic stability condition,
$\left(\frac{\partial P_{\rm eff}}{\partial V}\right)_S<0$, requires $n<0$.
Thus, unlike the small-volume regime, where the stability behaviour is primarily governed by $\xi_0$, the late-time stability is controlled by the sign of $n$. This behaviour is also consistent with the result shown in Fig.~\ref{pv}.
\end{enumerate}

Next, we discuss specialized scenarios to define the limits of these parameters.\\
\begin{enumerate}[label=(\roman*)]
\item The special case $n=0$, $\alpha=0$, and $\xi_0=0$ is excluded from the stability analysis, as it leads to a marginal thermodynamic response.
For these parameter values,
$\left(\frac{\partial P_{\rm eff}}{\partial V}\right)_S=0$,
indicating that the pressure remains unchanged under an adiabatic
variation of the volume. In this limit, $B_0$ effectively plays the
role of a cosmological constant, and the corresponding cosmological
solution reduces to a de Sitter-type spacetime.

\item  For $\alpha=0$ and $\xi_0=0$, with $n\neq0$, Eq.~\eqref{eq:eq18} takes the form
$\left(\frac{\partial P_{\rm eff}}{\partial V}\right)_S
= \frac{n}{3}B_0V^{-\left(1+\frac{n}{3}\right)}$.
Thus, for $n<0$, we obtain
$\left(\frac{\partial P_{\rm eff}}{\partial V}\right)_S<0$,
indicating thermodynamic stability. This result differs from that of
Santos \emph{et al.}~\cite{san1}, who obtained
$\left(\frac{\partial P_{\rm eff}}{\partial V}\right)_S=0$ for
$\alpha=0$. Furthermore, when $n\leq0$, the pressure becomes
increasingly negative as the volume increases, in agreement with the
observational results reported in Ref.~\cite{guo1}.

\item When $\xi_0 = 0$, $n = 0$ and $ \alpha \neq 0$, the  Eq. \eqref{eq:eq27}
 reduces to Santos's work
~\cite{san1}. In this case   $ \left( \frac{\partial P_{\rm eff}}
{\partial V} \right)_S$ will be

\begin{eqnarray}\label{eq:eq28}
 \left( \frac{\partial P_{\rm eff}}{\partial
V}\right)_{S}  =  \alpha \frac{ P_{\rm eff}}{V}
\left(\frac{\epsilon}{V} \right)^{1+\alpha} \left \{ 1+ \left(\frac{\epsilon}{V}
 \right )^{1 + \alpha} \right\}^{ - 1},
 \end{eqnarray}
 which gives  $ \left( \frac{\partial P_{\rm eff}}{\partial
V}\right)_{S}   < 0 $ for $\alpha >0$. It also agrees with the work of Santos
\emph{et al}~\cite{san1} in this field.

\item  when  $\alpha = 0$, $n = 0$ and $ \xi_0 \neq 0$,  $\left( \frac{\partial P_{\rm eff}}{\partial
V}\right)_{S} =  \frac{ \sqrt{3}\xi_0 B_0}{V} \left(\frac{\epsilon}{V} \right)$, \emph{i.e.},
$\left( \frac{\partial P_{\rm eff}}{\partial
V}\right)_{S}  > 0 $
which is positive for $\xi_0>0$ and $B_0>0$. In contrast, for $\xi_0<0$, the same derivative becomes negative, provided $B_0>0$. Therefore, the adiabatic stability condition $\left(\frac{\partial P{\rm eff}}{\partial V}\right)_S<0$ is satisfied only for negative $\xi_0$ in this special case. This further supports the preference for the $\xi_0<0$ parameter regime obtained from the thermodynamic stability analysis.

\item Now for $\alpha =1$, $\xi_0 \neq 0$ \& $n \neq 0$ we get from \eqref{eq:eq27} reduces to

\begin{eqnarray}\label{eq:eq29a}
\left( \frac{\partial P_{\rm eff}}{\partial V}\right)_{S}
=\frac{P_{\rm eff}}{6V}\frac{ - 2\sqrt{3}\xi_0  \left\{n \left[1 +
\left(\frac{\epsilon}{V} \right)^N \right]
 + 3N\left(\frac{\epsilon}{V} \right)^N \right \}
  + N \left\{\frac{3N \alpha \left(\frac{\epsilon}{V} \right)^N}{\left[
1+ \left(\frac{\epsilon}{V} \right)^N \right]} - n  \right\}  }{N+ 2 \sqrt{3}\xi_0
\left[1 +
\left(\frac{\epsilon}{V} \right)^N \right]}.~
\end{eqnarray}
Thus, for $\xi_0<0$ and $n<0$, we obtain $\left(\frac{\partial P_{\rm eff}}{\partial V}\right)_S<0$, indicating that the thermodynamic stability condition is satisfied. This finding is also consistent with the results obtained by Barman and Barman~\cite{bar}.

\item Again when  $\xi_0 =0$ but $\alpha \neq 0$ \& $n \neq 0$ the Eq. \eqref{eq:eq27} gives

\begin{eqnarray}\label{eq:eq30}
\left( \frac{\partial P_{\rm eff}}{\partial V}\right)_{S} =  \frac{P_{\rm eff}}{3V (1+\alpha)}
\left[ \frac{3N \alpha\left( \frac{\epsilon }{V}\right)^N}{  1 +  \left(\frac
{\epsilon }{V}\right)^N } - n \right].
\end{eqnarray}
This corresponds to the Variable generalised Chaplygin gas (VGCG) model~\cite{dp3, dp4}. It is clear from Eq.~\eqref{eq:eq30} that, for $\alpha>0$, $n$ must be negative for $\left(\frac{\partial P_{\rm eff}}{\partial V}\right)_S<0$ to be satisfied throughout the evolution. In this context, Lu~\cite{lu} investigated the Variable Generalised Chaplygin Gas (VGCG) model using the Union SNe Ia data and the Sloan Digital Sky Survey (SDSS) baryon acoustic peak and obtained the best-fit values $n=0.75$ and $\alpha=1.53$. However, these parameter values do not satisfy the thermodynamic stability condition obtained in the present analysis.
It is worth noting that, in an earlier article~\cite{dp1}, the case $ \xi_0 = 0 $  with  $ \alpha = 1$  and  $n \neq 0 $ was considered, and it was found that $ \left(\frac{\partial P}{\partial V}\right)_S < 0 $ is satisfied throughout the evolution for $ n < 0$. These results further support the requirement of negative $n$ for thermodynamic stability.
\end{enumerate}
However, our analysis involves multiple parameters: $B_0$, $\alpha$, $\xi_0$, and $n$. Since constraining all parameters simultaneously from Eq.~\eqref{eq:eq27} is not possible, we adopt a step-by-step approach. Here, we focus specifically on the roles of $n$ and $\xi_0$ rather than the other parameters, as they govern the key evolutionary features of the viscous variable generalised Chaplygin gas (VVGCG) model.

It follows from Eq. \eqref{eq:eq27} that, for positive values of $B_0$ and $\alpha$, the condition $\left(\frac{\partial P_{\rm eff}}{\partial V}\right)_S < 0$ holds when both $n < 0$ and $\xi_0 < 0$ throughout the evolution. While the influence of $\xi_0$ is dominant initially, $n$ plays a crucial role in driving acceleration in the later universe. Specifically, for $\xi_0<0$ and $n<0$, the inequality $\left(\frac{\partial P_{\rm eff}}{\partial V}\right)_S<0$ is satisfied throughout the cosmic evolution. Therefore, from the standpoint of thermodynamic stability, we conclude that both $n$ and $\xi_0$ must be negative for the VVGCG model.

It is to be mentioned that the behavior depicted in Fig.~\ref{dpv1} leads to several key insights. As the volume ($V$) increases, the derivative $\left(\frac{\partial P_{\rm eff}}{\partial V}\right)_S$ remains strictly negative across the entire domain, meaning the effective pressure ($P_{\rm eff}$) decreases (becomes more negative) to maintain mechanical stability. In the small-volume regime, the curve drops sharply toward large negative values, where the negative parameter $\xi_0 < 0$ governs and amplifies this steep response, ensuring a smooth, continuous transition free of unphysical singularities. As the volume expands toward larger scales, the curve rises monotonically and flattens out, gradually approaching zero from below to maintain a negative slope. Ultimately, this behaviour confirms that the combined negative values of $n$ and $\xi_0$ sustain a stable evolutionary branch within the viscous variable generalized Chaplygin gas (VVGCG) framework.

Next, we must examine whether the condition $\left(\frac{\partial P}{\partial V}\right)_T \le 0$ is also satisfied. We will demonstrate in the next section that this condition can be fulfilled for $n < 0$.

\subsection{Thermal capacity at Constant Volume ($c_V$)}
Another important criterion for thermodynamic stability is the positivity of the heat capacity at constant volume, $c_V$, which is given by $ c_V=T\left(\frac{\partial S}{\partial T}\right)_V
=\left(\frac{\partial U}{\partial T}\right)_V
=V\left(\frac{\partial \rho}{\partial T}\right)_V $.
To examine this stability condition, we first determine the temperature $T$ of the viscous variable generalised Chaplygin gas as a function of the volume $V$ and entropy $S$. The temperature follows from the thermodynamic relation
$T=\left(\frac{\partial U}{\partial S}\right)_V $.
Using this relation, together with Eq.~\eqref{eq:eq9}, we obtain the following expression for the temperature $T$:

\begin{subequations}\label{eq:eq31}
\begin{align}
T &=
\frac{V^{1-N-\frac{n}{3(1+\alpha)}}}{1+\alpha}\left[\frac{B_{0}(1+\alpha)}{N}
+ V^{-N}c \right]^{-\frac{\alpha}{1+\alpha}}\left( \frac{\partial c}{\partial S} \right)_{V}   \label{eq:eq31a} \\
&= \frac{V^{1-N-\frac{n}{3}}}{1+\alpha}\rho^{- \alpha} \left( \frac{\partial c}{\partial S} \right)_{V}  \label{eq:eq31b}.
\end{align}
\end{subequations}
If $c$ is further assumed to be a universal constant, then $ \frac{dc}{dS}=0$, implying that the fluid remains at zero temperature for all values of its volume and pressure. Therefore, to examine the thermodynamic stability of the viscous variable generalised Chaplygin gas, whose temperature is expected to evolve during expansion, we require the derivative of Eq.~\eqref{eq:eq31a} with respect to $S$ to be non-zero, i.e., $\left(\frac{\partial c}{\partial S}\right)\neq 0$. Since there is no \emph{a priori} prescription for the dependence of $c$ on $S$, its functional form cannot be determined uniquely. Nevertheless, physical considerations require the temperature to remain positive and to decrease during adiabatic expansion. Accordingly, we impose the condition  $\left(\frac{\partial c}{\partial S}\right) > 0$. From dimensional analysis of Eq.~\eqref{eq:eq9}, we further obtain

\begin{equation}\label{eq:eq32}
 \left[U \right] = \left\{\frac{\left[c \right]}{\left[V \right
 ]^{{-\sqrt{3}\xi_0}(1+\alpha)}} \right\}^{\frac{1}{1+ \alpha}}.
\end{equation}
Since $\left[U \right] = \left[T\right]\left[S \right]$,  we can write

 \begin{equation}\label{eq:eq33}
 \left[c \right] = \left[T \right]^{1+ \alpha}\left[S \right]^{1+ \alpha}\left[V \right]^{{-\sqrt{3}\xi_0}(1+ \alpha)}.
\end{equation}
Since an analytic solution for $c$ from Eq. \eqref{eq:eq33} is difficult to obtain, we consider, as a trial case, an empirical form for $c$ and investigate whether the resulting expressions are consistent with the standard thermodynamic relations. As $c$ is assumed to depend solely on the entropy, we adopt the following form:

\begin{equation}\label{eq:eq34}
 c = \left( \tau v^{-\sqrt{3}\xi_0} \right)^{1+ \alpha} S^{1+ \alpha},
\end{equation}
where \(\tau\) and \(v\) are constants with dimensions of time and volume, respectively. Now

\begin{equation}\label{eq:eq35}
\frac{dc}{dS} = (1+ \alpha) \left( \tau v^{-\sqrt{3}\xi_0} \right)^{1+ \alpha} S^{\alpha}.
\end{equation}
Using Eqs. \eqref{eq:eq31a} and \eqref{eq:eq35}, we get the expression of temperature

\begin{subequations}\label{eq:eq36}
\begin{align}
T &=
V^{1-N-\frac{n}{3(1+\alpha)}}\left[\frac{B_{0}(1+\alpha)}{N}
+ V^{-N}c \right]^{-\frac{\alpha}{1+\alpha}}\left( \tau v^{-\sqrt{3}\xi_0} \right)^{1+\alpha}S^{\alpha}   \label{eq:eq36a} \\
&= V^{1-N-\frac{n}{3}} \left( \tau v^{-\sqrt{3}\xi_0} \right)^{1+ \alpha} \rho^{- \alpha} S^{\alpha} \label{eq:eq36b}\\
&=\frac{\tau v^{-\sqrt{3}\xi_0}}{V^{-\sqrt{3}\xi_0}} \left \{ 1- \frac{1}{1+ \left(\frac{\epsilon}{V} \right)^N} \right \}^{\frac{\alpha}{1+ \alpha}},   \label{eq:eq36c}
\end{align}
\end{subequations}
and from Eq. \eqref{eq:eq36a},  the entropy is

\begin{equation}\label{eq:eq37}
S = \frac{\left[\frac{B_{0}(1+\alpha)}{N} V^{N} \right]^{\frac{1}{1+\alpha}} \left(\frac{T}{\tau^{1+\alpha}} \right)^{\frac{1}{\alpha}} \left(\frac{V}{v^{1+\alpha)}} \right)^{\frac{-\sqrt{3} \xi_0}{\alpha}}}{\left \{1-\left(\frac{T}{\tau} \right)^{\frac{1+\alpha}{\alpha}} \left(\frac{V}{v} \right)^{\frac{-\sqrt{3}\xi_0(1+\alpha)}{\alpha}}\right \}^{\frac{1}{1+\alpha}}}~,
\end{equation}
Equation \eqref{eq:eq37} indicates that, for the entropy to remain positive and finite, the condition $ 0 < TV^{-\sqrt{3}\xi_0} < \tau v^{-\sqrt{3}\xi_0}
$ must be satisfied, together with the individual bounds
$ 0<T<\tau$ \& $v<V<\infty$. Here, $ \tau $ and $v$ denote the maximum temperature and minimum volume, respectively. Furthermore, Eq. \eqref{eq:eq36c} shows that the temperature of the VVGCG decreases monotonically with increasing volume for $\xi_0 < 0$. In particular, as the volume approaches its minimum value, $V\rightarrow v$, the temperature approaches its maximum value, $T\rightarrow\tau$, whereas in the large-volume limit, $ V\rightarrow\infty$, the temperature tends to zero, $T\rightarrow0$. Thus, the thermodynamic evolution is restricted to a finite minimum volume and a finite maximum temperature, rather than extending to arbitrarily small volumes.

Evidently, as $T\rightarrow0$, the entropy vanishes, indicating that the third law of thermodynamics is satisfied in this case.
Now using Eq. \eqref{eq:eq37}
 we get the expression of thermal heat capacity as

\begin{equation}\label{eq:eq38}
c_{V}  = T\left(\frac{\partial S}{\partial T} \right)_{V} =\frac{\left[\frac{B_{0}(1+\alpha)}{N} V^{N} \right]^{\frac{1}{1+\alpha}} \left(\frac{T}{\tau^{1+\alpha}} \right)^{\frac{1}{\alpha}} \left(\frac{V}{v^{1+\alpha}} \right)^{\frac{ - \sqrt{3}\xi_0}{\alpha}}}{ \alpha \left \{1-\left(\frac{T}{\tau} \right)^{\frac{1+\alpha}{\alpha}}\left(\frac{V}{v} \right)^{\frac{- \sqrt{3}\xi_0(1+\alpha)}{\alpha}}\right \}^{\frac{2+ \alpha}{1+\alpha}}}~.
\end{equation}

Since $0 < TV^{-\sqrt{3}\xi_0} < \tau v^{-\sqrt{3}\xi_0}$ and $\alpha > 0$, the condition $c_V > 0$ is consistently satisfied regardless of the values of $n$ and $\xi_0$. It is noteworthy that as the temperature approaches zero, $c_V$ vanishes, in accordance with the third law of thermodynamics. Moreover, setting $\xi_0 = 0$ and $\alpha = 1$ recovers the Variable Chaplygin gas model, yielding an expression identical to that in our previous work~\cite{dp2}. Finally, for $\xi_0 = 0$ and $n = 0$, Eq. \eqref{eq:eq38} reduces to the formulation presented by Santos et al.~\cite{san1}.

To conclude this section, a final remark is in order. While the positivity of specific heat is strongly desirable in the context of special relativity, Luongo \emph{et al.}~\cite{luo} recently argued that in an FLRW-type model like the one discussed here, a negative specific heat at constant volume and a vanishingly small specific heat at constant pressure ($c_P$) are compatible with observational data. In fact, they derived the most general cosmological model consistent with the values $c_V < 0$ and $c_P \sim 0$ obtained for the universe, showing in addition that it overcomes both the fine-tuning and coincidence problems of the $\Lambda$CDM model.

\subsection{Thermal equation of state}\label{tes}

As the pressure depends on both volume and temperature, Eqs. \eqref{eq:eq9}, \eqref{eq:eq34}, and \eqref{eq:eq37} yield the internal energy as a function of $V$ and  $T$ in the following form:

\begin{equation}\label{eq:eq39}
  U= V \left \{ \frac{\frac{B_{0}(1+\alpha)}{N} V^{-\frac{n}{3}}}{ {  1-\left(\frac{T}{\tau} \right)^{\frac{1+\alpha}{\alpha}}\left(\frac{V}{v} \right)^{\frac{ - \sqrt{3}\xi_0(1+\alpha)}{\alpha}}}} \right\}^{\frac{1}{1+ \alpha}}.
\end{equation}
Using Eqs. \eqref{eq:eq5}, \eqref{eq:eq6}, and \eqref{eq:eq39}, the pressure can be expressed as
\begin{subequations}\label{eq:eq40}
\begin{align}
P_{\rm eff} & = -\sqrt{3} \xi_0 - \left(B_0 V^{-\frac{n}{3}} \right)^{\frac{1}{1 + \alpha}} \left( \frac{N}{1+\alpha} \right)^{\frac{\alpha}{1+\alpha}}
\left \{1- \left(\frac{T}{\tau} \right)^{\frac{1+\alpha}{\alpha}} \left(\frac{V}{v} \right)^{-\frac{\sqrt{3} \xi_0 (1+\alpha)}{\alpha}} \right \}^{\frac{\alpha}{1+\alpha}}  \label{eq:eq40a}\\
&= \rho_{\rm vvgcg} \left[ - \sqrt{3}\xi_0 - \frac{N}{1+ \alpha}  \left \{{  1-\left(\frac{T}{\tau} \right)^{\frac{1+\alpha}{\alpha}}\left(\frac{V}{v} \right)^{\frac{ - \sqrt{3}\xi_0(1+\alpha)}{\alpha}}} \right \} \right ]. \label{eq:eq40b}
\end{align}
\end{subequations}
Eqs \eqref{eq:eq40a} \& \eqref{eq:eq40b} shows that, for $\xi_0>0$, $P_{\rm eff}<0$ throughout the evolution, whereas for $\xi_0<0$, $P_{\rm eff}$ is initially positive, vanishes at a critical volume, and subsequently becomes negative with increasing volume, consistent with the behavior discussed in Sec. \ref{pr}.  Eq \eqref{eq:eq40a} shows that, for $\xi_0=0$ and $\alpha=1$, the solution reduces to an earlier work~\cite{dp1}. Furthermore, in the case $\xi_0=0$ and $n=0$, the above solution reduces to the result obtained by Santos et al.~\cite{san1}. Now the thermal EoS parameter is then given by

\begin{equation}\label{eq:eq41}
 \omega = - \sqrt{3}\xi_0 - \frac{N}{1+ \alpha}  \left \{{  1-\left(\frac{T}{\tau} \right)^{\frac{1+\alpha}{\alpha}}\left(\frac{V}{v} \right)^{\frac{ - \sqrt{3}\xi_0(1+\alpha)}{\alpha}}} \right \}.
\end{equation}
Thus, the thermal EoS parameter is, in general, a function of both temperature and volume. In the high-temperature, small-volume limit, \emph{i.e.}, as $T\rightarrow\tau$ and $V\rightarrow v$, Eq.~\eqref{eq:eq41} gives $\omega\approx-\sqrt{3}\xi_0$. Consequently, $P_{\rm eff}\approx-\sqrt{3}\xi_0\rho_{\rm vvgcg}$, which coincides with Eq.~\eqref{eq:eq19}. Therefore, for $\xi_0<0$, $P_{\rm eff}>0$ in this regime, consistent with the behavior discussed in Sec. \ref{pr}.

Secondly, in the large-volume limit, corresponding to a very low temperature, \emph{i.e.}, as $T \rightarrow 0$, Eq. \eqref{eq:eq41} yields $\omega \approx -1 + \frac{n}{3(1+\alpha)}$. This expression is identical to Eq. \eqref{eq:eq20}, consistent with the results reported in the existing literature~\cite{san1,dp2}. Therefore, the thermodynamic states described by Eqs. \eqref{eq:eq19} and \eqref{eq:eq20} are recovered in the respective early- and late-time limits of the Universe.

\subsection{Isothermal Stability Condition}\label{isc}

We now examine the isothermal stability condition,
$\left(\frac{\partial P_{\rm eff}}{\partial V}\right)_{T} \leq 0$.
Starting from Eq.~\eqref{eq:eq40a}, we obtain
\begin{equation}\label{eq:eq42}
\begin{split}
\left( \frac{\partial P_{\rm eff}}{\partial V} \right)_T & = \frac{\left(B_0 V^{-\frac{n}{3}} \right)^{\frac{1}{1 + \alpha}} \left( \frac{N}{1+\alpha} \right)^{\frac{\alpha}{1+\alpha}}}{V \left \{{  1-\left(\frac{T}{\tau} \right)^{\frac{1+\alpha}{\alpha}}\left(\frac{V}{v} \right)^{\frac{ - \sqrt{3}\xi_0(1+\alpha)}{\alpha}}} \right \} } \\
& \left[  \frac{n}{3(1+\alpha)} \left \{{  1-\left(\frac{T}{\tau} \right)^{\frac{1+\alpha}{\alpha}}\left(\frac{V}{v} \right)^{\frac{ - \sqrt{3}\xi_0(1+\alpha)}{\alpha}}} \right \}
 +\sqrt{3} \xi_0  \left(\frac{T}{\tau} \right)^{\frac{1+\alpha}{\alpha}}\left(\frac{V}{v} \right)^{\frac{ - \sqrt{3}\xi_0(1+\alpha)} {\alpha}} \right]
\end{split}
\end{equation}
Equation~\eqref{eq:eq42} shows that, for negative values of $n$ and $\xi_0$, the condition $\left(\frac{\partial P_{\rm eff}}{\partial V}\right)_T<0$ is satisfied throughout the evolution.

The thermodynamic stability of the VVGCG requires the fulfillment of both the isothermal and adiabatic stability conditions, namely, $\left(\frac{\partial P_{\rm eff}}{\partial V}\right)_T<0$ and $\left(\frac{\partial P_{\rm eff}}{\partial V}\right)_S<0$, respectively. Our analysis demonstrates that both conditions are satisfied for negative values of $n$ and $\xi_0$.

It is also important to distinguish our treatment from that of Santos et al.~\cite{san2}, who assumed $\left(\frac{\partial P_{\rm eff}}{\partial V}\right)_T=0$ as a simplifying condition. In the present analysis, we retain the full isothermal dependence of the pressure and determine the function $c=c(S)$ through an alternative approach. This allows us to consider a more general thermodynamic framework without imposing the aforementioned simplifying assumption. Consequently, the present formulation does not reduce to the corresponding case considered by Santos et al.~\cite{san2} in the limit $n=0$.\\
\begin{figure}[t]
\begin{center}
   \includegraphics[width=8cm]{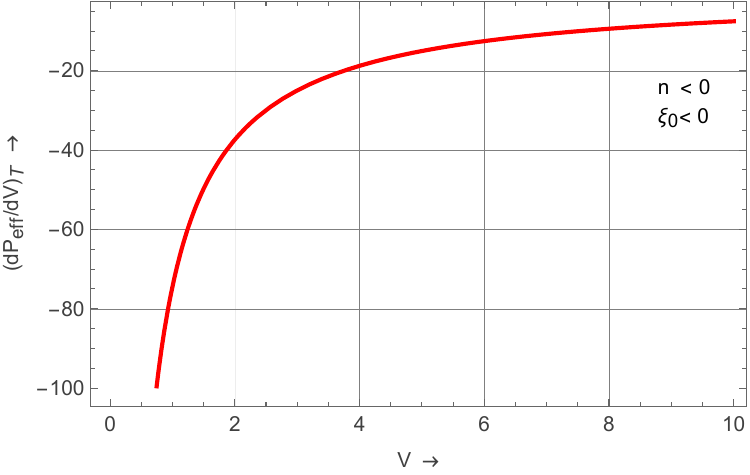}
     \caption{
  \small\emph{ The variations of $\left(\frac{\partial P_{\rm eff}}{\partial V}\right)_{T}$ vs
   $V$   are shown. The graphs clearly show that $\left(\frac{\partial P_{\rm eff}}{\partial V}\right)_{T} <0$ throughout the evolution only for negative values of $n$ \& $\xi_0$.
          }\label{dpvt}
    }
\end{center}
\end{figure}
To investigate the thermodynamic stability of the viscous Variable Generalized Chaplygin Gas (VVGCG) model, we examine the isothermal derivative $(\partial P_{\rm eff}/\partial V)_T$. For $\xi_0<0$, this derivative remains negative throughout the considered volume range, as shown in Fig.~\ref{dpvt}, indicating that the isothermal stability condition is satisfied. It remains well behaved from the small-volume regime, where the minimum volume scale $v$ prevents singular behavior, to the large-volume regime corresponding to late-time cosmic evolution.

The negativity of $(\partial P_{\rm eff}/\partial V)_T$ implies a positive isothermal compressibility and hence ensures isothermal mechanical stability. Thus, $\xi_0<0$ corresponds to a thermodynamically viable branch of the VVGCG model, consistent with the stability requirements established in this work.

\subsection{Consistency of the Thermodynamic Relations}\label{ctr}
In the preceding Sec \ref{tes}, we derived the expression for the temperature using dimensional analysis based on Eq.~\eqref{eq:eq34} and subsequently investigated the entropy, heat capacity at constant volume,  and other thermodynamic quantities. However, the consistency of these expressions with the complete set of thermodynamic relations cannot be established solely from their individual derivations. It is therefore necessary to subject the solutions obtained from Eq.~\eqref{eq:eq34} to an independent thermodynamic consistency check. To this end, we consider the standard thermodynamic relation for the internal energy, namely the first internal-energy equation~\cite{zym}, as a test case. In particular, we examine whether the expression for the internal energy given by Eq.~\eqref{eq:eq34} satisfies the following thermodynamic identity:

\begin{equation}\label{eq:eq43}
 \left(\frac{\partial U}{\partial V} \right)_{T} = T  \left(\frac{\partial P_{\rm eff}}{\partial T} \right)_{V} -P_{\rm eff}.
\end{equation}
Using Eqs \eqref{eq:eq6} and \eqref{eq:eq39} we get the LHS of Eq \eqref{eq:eq43} as

\begin{equation}\label{eq:eq44}
 \left(\frac{\partial U}{\partial V} \right)_{T} = \rho_{\rm vvgcg} \left[  1 - \frac{n}{1+\alpha} - \frac{\sqrt{3}\xi_0   \left(\frac{T}{\tau} \right)^{\frac{1+\alpha}{\alpha}}\left(\frac{V}{v} \right)^{\frac{ - \sqrt{3}\xi_0(1+\alpha)} {\alpha}}  }  {\alpha \left \{ 1-  \left(\frac{T}{\tau} \right)^{\frac{1+\alpha}{\alpha}}\left(\frac{V}{v} \right)^{\frac{ - \sqrt{3}\xi_0(1+\alpha)} {\alpha}} \right \}} \right]
\end{equation}

Evaluating the temperature derivative via Eq.~\eqref{eq:eq40b}, we obtain
\begin{equation}\label{eq:eq44a}
 T  \left(\frac{\partial P_{\rm eff}}{\partial T} \right)_{V}   = \rho_{\rm vvgcg}  \left(\frac{T}{\tau} \right)^{\frac{1+\alpha}{\alpha}} \left(\frac{V}{v} \right)^{\frac{ - \sqrt{3}\xi_0(1+\alpha)} {\alpha}}\left[   \frac{N}{1 + \alpha} -  \frac{\sqrt{3}\xi_0   }{  \alpha \left \{ 1- \left(\frac{T}{\tau} \right)^{\frac{1+\alpha}{\alpha}}\left(\frac{V}{v} \right)^{\frac{ - \sqrt{3}\xi_0(1+\alpha)} {\alpha}}  \right \}}  \right]
\end{equation}
\noindent Subsequently, substituting Eqs.~\eqref{eq:eq40b} and \eqref{eq:eq44a}, we determine the expression for the right-hand side (RHS) of Eq.~\eqref{eq:eq43}
\begin{equation}\label{eq:eq45}
 T  \left(\frac{\partial P_{\rm eff}}{\partial T} \right)_{V} - P_{\rm eff}  = \rho_{\rm vvgcg} \left[  -  \frac{\sqrt{3}\xi_0   \left(\frac{T}{\tau} \right)^{\frac{1+\alpha}{\alpha}}\left(\frac{V}{v} \right)^{\frac{ - \sqrt{3}\xi_0(1+\alpha)} {\alpha}} }{  \alpha \left \{ 1- \left(\frac{T}{\tau} \right)^{\frac{1+\alpha}{\alpha}}\left(\frac{V}{v} \right)^{\frac{ - \sqrt{3}\xi_0(1+\alpha)} {\alpha}}  \right \}}    + \frac{N}{1 + \alpha} + \sqrt{3}\xi_0  \right]
\end{equation}
By comparing Eqs.~\eqref{eq:eq44} and \eqref{eq:eq45}, we obtain $N=(1-\sqrt{3}\xi_0)(1+\alpha)-\frac{n}{3}$, which is identical to the expression for $N$ given in Eq.~\eqref{eq:eq10a}.  This agreement provides an independent consistency check of the thermodynamic formulation adopted in our analysis and also confirms the consistency of our approach.

\section{ Observational constraints}\label{oc}

A comparison with observational data is essential to examine whether the proposed cosmological model provides a realistic description of the observed Universe. In particular, observational constraints on the model parameters allow us to test whether the theoretically motivated parameter ranges are compatible with the cosmological evolution inferred from observations. Therefore, consistency with the Hubble-$57$ data provides an important test of the viability of the VVGCG model and helps identify the observationally favoured region of its parameter space.

In this section, we use the Hubble-$57$ data set~\cite{dp5} to constrain the model parameters and assess the consistency of the VVGCG model with observations. The Hubble parameter $H(z)$ at a given redshift $z$ can be determined using two main approaches: (i) estimation from the differential ages (DA), $\Delta t$, of galaxies; and (ii) extraction from line-of-sight BAO measurements, including the analysis of the correlation functions of luminous red galaxies. The detailed data table has been presented in our recent article~\cite{dp3}. The Hubble parameter obtained using the differential-age method can be expressed as a function of the redshift $z$ in the form

\begin{equation}\label{eq:46}
H(z) = -\frac{1}{1+z} \frac{dz}{dt} \simeq  -\frac{1}{1+z} \frac{\vartriangle z}{\vartriangle t}
\end{equation}
therefore,  $H(z)$ can be found directly from Eq.~\eqref{eq:46} once $\frac{dz}{dt}$ is known ~\cite{sei}. Considering $V = a^3$ and using the present value of the scale factor  normalised to unity , \emph{i.e.}, $a = a_0 =1$ we get a relation of the Hubble parameter with the redshift parameter $z$. If $\rho_0$ be the density at present epoch  then the well known density parameter was written previously as   $ \Omega_x = \frac{(1+\alpha)B_0}{N\,\rho_0^{\,1+\alpha}} $ ~\cite{seth} which yields $c =\rho_0^{1+\alpha} \left(1- \Omega_x \right)$. Now using eq.~\eqref{eq:eq8}, we can express the three space matter density as

\begin{equation}\label{eq:eq47}
\rho_{vvgcg} =  \rho_{0} (1+z)^{\frac{n}{1+\alpha}}\left\{ \Omega_x + (1- \Omega_x) (1+z)^{3(1+\alpha) \left(1-\sqrt{3}\xi_0 \right) - n }\right\}^{\frac{1}{1+\alpha}}
\end{equation}
From eqn~ \eqref{eq:eq47} we get
\vspace{-0.3 cm}
\begin{equation}\label{eq:eq48}
  H =  H_{0} (1+z)^{\frac{n}{2(1+\alpha)}}\left\{
\Omega_x + (1- \Omega_x) (1+z)^{3(1+\alpha) \left(1-\sqrt{3}\xi_0 )- n\right)}\right\}^{\frac{1}{2(1+\alpha)}}
\end{equation}
\vspace{-0.1 cm}
where $H_0 = \left(\frac{ \rho_0}{3} \right)^{\frac{1}{2}}$ is the present value of the Hubble parameter.
Equation~\eqref{eq:eq48} gives the redshift evolution of the Hubble parameter $H(z)$.

\begin{figure}[h!]
    \centering 
    \begin{tabular}{l l}
        \parbox{2in}{%
            \includegraphics[width=2in,height=2.1in]{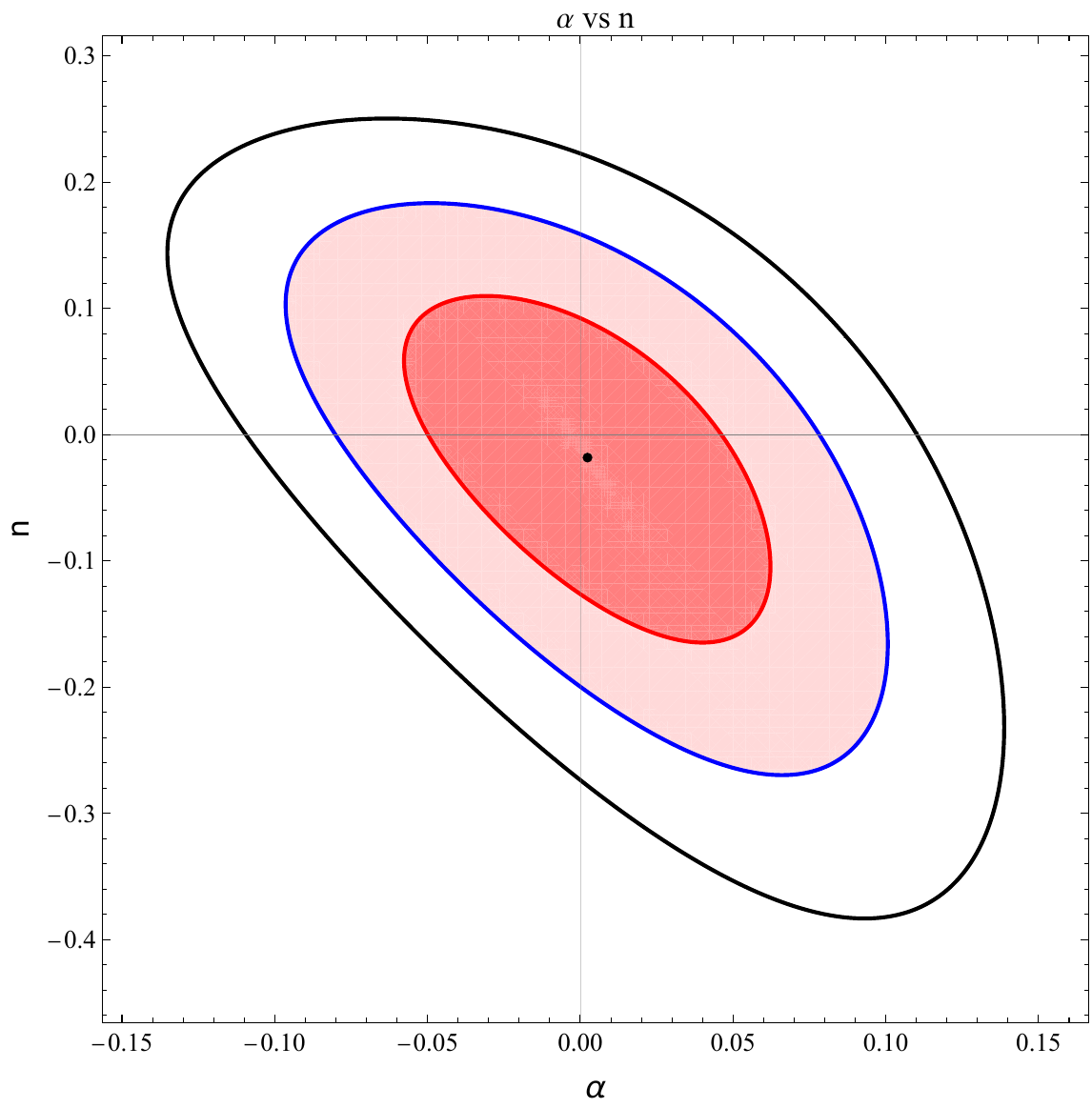}%

        } &
        \parbox{2in}{%
           \includegraphics[width=2in,height=1in]{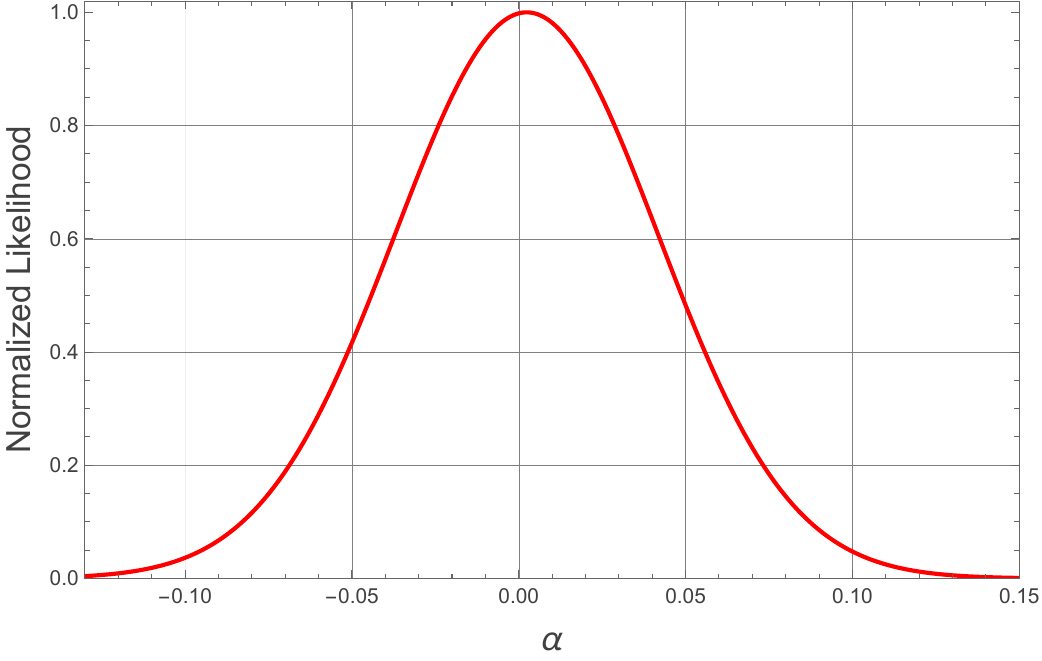}\\
            \includegraphics[width=2in,height=1in]{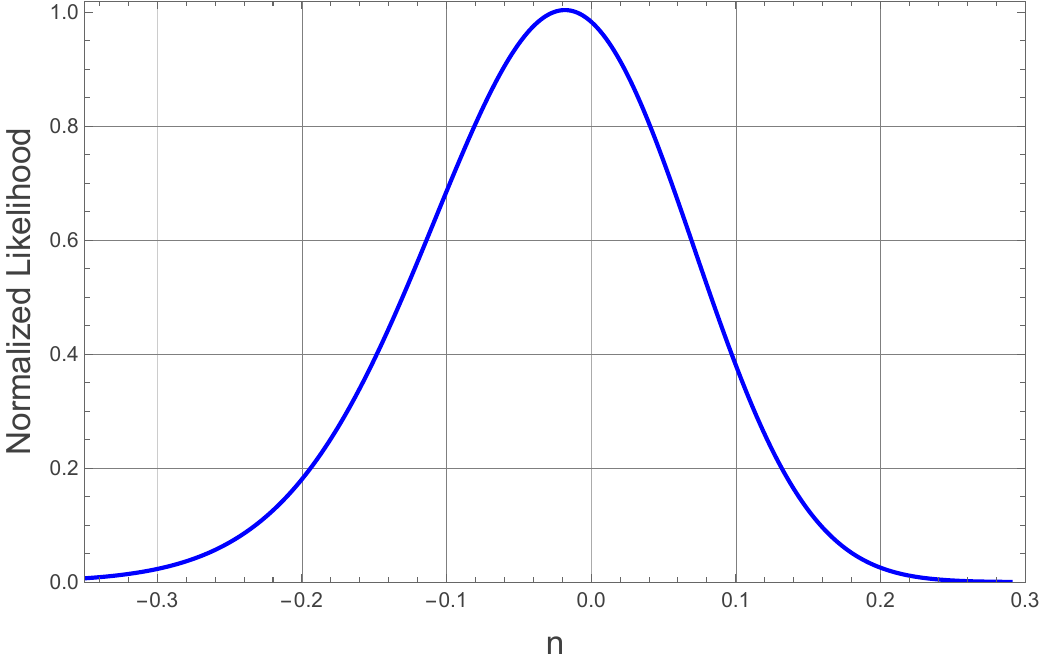}%
        } \\
    \end{tabular}
    \caption{\small\emph{$\xi_0$ vs $n$ graph with liklihood}} 
    \label{Lbsa}
\end{figure}

The best-fit values of the model parameters extracted from the contour plots are summarized in Table~\ref{tab:contour}.

\begin{table}[h]
\centering
\begin{tabular}{c c c c c c}
\hline
$ \chi^2_{min}$ & $H_0$ & $\Omega_x$ & $\xi_0$& $\alpha$ & $n$  \\
\hline
\vspace{0.1 cm}
$ 44.56$ &\hspace{0.1 cm} $70.8$ &\hspace{0.2 cm} $0.748$ & - 0.0015  & $  0.0024^{ +0 .062}_{-0.056} $ & $  -0.0182^{ + 0.10}_{-0.16}$ \\
\hline
\vspace{0.1 cm}
\end{tabular}
\caption{Best-fit values of the model parameters obtained from the contour plot analysis.}
\label{tab:contour}
\end{table}
The best-fit values of the model parameters obtained from the contour-plot analysis (Table~\ref{tab:contour}) reveal several important features of the VVGCG model. For the chosen values of $H_0$, $\Omega_x$, and a negative $\xi_0$, the corresponding best-fit values of the remaining parameters are $\alpha=0.0024$ and $n=-0.0182$. As discussed analytically in the preceding sections, thermodynamic
stability of the VVGCG fluid requires $\alpha>0$, $n<0$, and $k_0<0$
throughout its evolution. The obtained parameter values satisfy these
requirements and are therefore consistent with the thermodynamic
stability analysis and the constraints derived earlier. In particular, the small best-fit value $\alpha=0.0024$ is consistent with the result reported in our earlier communication, where it was found that $\alpha$ should be smaller in the late Universe~\cite{dp5}. Furthermore, the negative values of $n$ and $\xi_0$ play an important role in maintaining thermodynamic stability, particularly through the positivity of entropy and the behaviour of the sound speed. Thus, the observationally favoured parameter region naturally coincides with the thermodynamically stable branch of the model, providing further support for the physical viability of the VVGCG framework.

\section{ Discussion}\label{dis}

While many dark energy models can provide a satisfactory description of cosmological observations, their thermodynamic consistency and stability require further examination. In this work, we investigate the cosmological dynamics and thermodynamic stability of a viscous variable generalized Chaplygin gas (VVGCG) model. In a homogeneous and isotropic FLRW spacetime, the bulk-viscous pressure is given by
$\Pi=-3\xi H$.
In standard linear irreversible thermodynamics, the bulk-viscosity coefficient is generally required to satisfy $\xi \geq0$ to ensure non-negative entropy production \cite{ons1, ons2}. However, effective descriptions of non-linear and far-from-equilibrium systems can admit a negative viscosity parameter under appropriate assumptions \cite{pri}. For $\xi<0$, the viscous pressure becomes positive, $\Pi>0$, thereby counteracting the negative equilibrium pressure of the Chaplygin gas. This can produce an early decelerating phase and facilitate a smooth transition to late-time cosmic acceleration.

Motivated by extensive investigations of viscous Chaplygin-type cosmological models \cite{saa, bre1, bre2}, we study the interplay between bulk-viscous effects and a variable generalized Chaplygin gas in a spatially flat FLRW universe. Following Ref.~\cite{tha}, we adopt the phenomenological dependence
$\xi=\xi_0\rho^{1/2}$
for the bulk-viscosity coefficient, where $\xi_0$ is a constant parameter whose sign plays a crucial role in determining the cosmic evolution and thermodynamic behavior of the model.

The main features of the model are summarized below:
\begin{enumerate}[label=(\roman*)]

\item  For $\xi_0<0$, the effective pressure is volume-dependent. It is positive, $P_{\rm eff}>0$, in the small-volume regime, vanishes at a finite critical volume $V_c$, and becomes negative, $P_{\rm eff}<0$, for $V>V_c$, as shown in Fig.~\ref{pv}. Thus, the model naturally describes a transition from an early decelerating phase to a late-time accelerating phase. In contrast, $\xi_0>0$ gives $P_{\rm eff}<0$ in the small-volume regime and hence maintains accelerated expansion from the outset. The critical volume $V_c$ is real and finite for $n<0$ and $\xi_0<0$. In the limit $\xi_0\rightarrow0$, the viscous contribution disappears and the transition associated with $V_c$ is absent, recovering the standard VGCG behavior.

\item  In the small-volume regime, $V\ll\epsilon$, Eq.~\eqref{eq:eq19} gives
$P_{\rm eff}\approx-\sqrt{3}\xi_0\rho_{\rm vvgcg}$.
Consequently, the sign of $\xi_0$ determines the early-time dynamics: $\xi_0>0$ leads to negative pressure and acceleration, whereas $\xi_0<0$ gives positive pressure and deceleration. For $\xi_0=0$, the viscous contribution vanishes and the fluid approaches dust-like behavior in this limit.

In the opposite regime, $V\gg\epsilon$, Eq.~\eqref{eq:eq20} gives
$\mathcal{W}\approx-1+\frac{n}{3(1+\alpha)}$.
For $0<n<3(1+\alpha)$, the asymptotic equation of state corresponds to a non-phantom accelerating regime; $n=0$ gives the $\Lambda$CDM limit, while $n<0$ leads to a phantom-like asymptotic state. The latter is compatible with observational analyses of phantom dark-energy scenarios \cite{cim}. As the volume increases, $\mathcal{W}$ evolves continuously from its early-time value to its asymptotic value, crossing $\mathcal{W}=0$ at $V=V_c$. When the thermodynamically stable branch $n<0$ and $\xi_0<0$ is considered, this behavior is illustrated in Fig.~\ref{wv}.

\item  From Eq.~\eqref{eq:eq18}, the deceleration parameter is given by Eq.~\eqref{eq:eq21}. In the small-volume limit, Eq.~\eqref{eq:eq22} shows that $\xi_0>0$ produces acceleration, whereas $\xi_0<0$ gives deceleration. At large volume, Eq.~\eqref{eq:eq23} yields $q<0$ for $n<0$, indicating late-time accelerated expansion.

The dynamical transition from deceleration to acceleration occurs at the flip volume $V_f$, defined by Eq.~\eqref{eq:eq24}. For $n<0$ and $\xi_0<0$, $V_f$ is real, as demonstrated in Fig.~\ref{qv}. The two characteristic volumes, $V_c$ and $V_f$, have distinct physical meanings: $V_c$ corresponds to $P_{\rm eff}=0$, whereas $V_f$ is determined by the condition $q=0$, or equivalently $\rho_{\rm vvgcg}+3P_{\rm eff}=0$ which follows from Eq.~\eqref{eq:eq21}. Since the onset of acceleration requires sufficiently negative pressure rather than merely $P_{\rm eff}<0$, the dynamical transition necessarily occurs after the pressure has crossed zero. Thus, the physically expected ordering is $V_c<V_f$, which is also obtained analytically in our model.

\item  We next examine the thermodynamic consistency and stability of the VVGCG fluid. Following the standard stability criteria,
$\left(\frac{\partial P_{\rm eff}}{\partial V}\right)_S < 0$,
$\left(\frac{\partial P {\rm eff}}{\partial V}\right)_T<0$, and
$c_V>0$,
we find that the thermodynamically stable branch requires $n<0$ and $\xi_0<0$, together with $\alpha>0$. These conditions are consistent with the analytical constraints obtained independently from the cosmological evolution.

The requirement $n<0$ leads to a phantom-like late-time equation of state. Phantom cosmologies have been investigated in connection with SNe~Ia and CMB observations \cite{cim}. In the present model, the same parameter choice that ensures thermodynamic stability also gives an early decelerating phase, followed by late-time phantom-like acceleration. Thus, the model provides a unified description connecting the early high-density regime with the asymptotic phantom regime. Whether the latter culminates in a finite-time Big-Rip singularity depends on the detailed asymptotic evolution of the scale factor and should therefore be established explicitly rather than assumed solely from $n<0$.

\item  The thermodynamic analysis further shows that the entropy vanishes at $T=0$, in accordance with the third law of thermodynamics. From the thermal and caloric equations of state, we obtain the physically allowed domain
$v<V<\infty,\qquad 0<T<\tau$,
where $v$ denotes the minimum attainable volume and $\tau$ the corresponding maximum temperature. In the limit $V\rightarrow v$, the temperature approaches its upper bound, $T\rightarrow\tau$. Thus, the model possesses a finite maximum temperature associated with the small but finite minimum volume of the fluid.

\item  Although the temperature, entropy, and heat capacity are derived from Eq.~\eqref{eq:eq34}, their individual expressions do not by themselves establish complete thermodynamic consistency. We therefore perform an independent consistency check using the internal-energy relation and verify whether it satisfies the standard thermodynamic identity given in Eq.~\eqref{eq:eq43} \cite{zym}.

\item Finally, we confront the model with the Hubble-57 data set \cite{dp5}. The resulting constraints, presented in Table~\ref{tab:contour}, yield the best-fit values $\alpha=0.0024$, $n=-0.0182$, and $\xi_0=-0.0015$. These fitted parameters satisfy the thermodynamic stability conditions $\alpha>0$, $n<0$, and $\xi_0<0$, ensuring thermodynamic stability throughout the evolution and consistency with the analytical bounds derived above.

\end{enumerate}
The small magnitude of $\alpha$ is consistent with its expected suppression in the late universe~\cite{dp5}. The constrained values of $n$ and $\xi_0$ further support the consistency of the observationally favoured parameter set with the theoretical requirements of the VVGCG model.

As a final remark, the framework is sufficiently general to recover several existing Chaplygin-gas models as limiting cases. It thus provides a unified description of cosmic evolution from an early decelerating phase to a late-time phantom-like regime while remaining thermodynamically stable throughout the evolution.

 \vspace{0.5cm}

\textbf{Acknowledgment : } We acknowledge Dr. S. Chatterjee for valuable suggestions.\\

\textbf{Declaration of competing interest}
The authors declare that they have no conflict of interest.

\end{document}